\documentclass[twocolumn,superscriptaddress,floatfix,preprintnumbers, nofootinbib,hyperref]{revtex4} 
\pdfoutput=1
\usepackage{hyperref}
\hypersetup{
    colorlinks=true,
    linkcolor=blue,
    filecolor=magenta,      
    urlcolor=cyan,
}
\usepackage{cancel}
\usepackage{ulem} 
\usepackage[utf8]{inputenc}
\hypersetup{allcolors=[rgb]{0.0 0.0 0.6},linkcolor=[rgb]{0.75 0.05 0.05}}
\usepackage{amsmath,amssymb}
\usepackage{epsfig}  
\usepackage{graphicx}   
\usepackage{slashed}             
\usepackage{url}
\usepackage{color}
\usepackage{multirow}

\usepackage[dvipsnames]{xcolor}
\usepackage{letltxmacro}
\LetLtxMacro{\oldcite}{\cite}
\renewcommand{\cite}[1]{\mbox{\oldcite{#1}}}

\DeclareMathOperator{\gcm}{g/cm^3}
\DeclareMathOperator{\cm}{cm}
\DeclareMathOperator{\GeV}{GeV}
\DeclareMathOperator{\eV}{eV}
\DeclareMathOperator{\meV}{meV}
\DeclareMathOperator{\keV}{keV}
\DeclareMathOperator{\MeV}{MeV}

\DeclareMathOperator{\s}{s}

\DeclareMathOperator{\erg}{erg}

\DeclareMathOperator{\km}{km}

\DeclareMathOperator{\kpc}{kpc}
\DeclareMathOperator{\g}{g}

\newcommand{\bp}{{\bf p}}

\newcommand{\beq}{\begin{equation}}
\newcommand{\eeq}{\end{equation}}
\newcommand{\nsat}{{n_{\rm sat}}}

\allowdisplaybreaks

\bibpunct{[}{]}{,}{n}{}{,}

\begin{document}
\preprint{INT-PUB-20-039}
\title{Enhanced Supernova Axion Emission and Its Implications} 

\author{Pierluca Carenza}
\email{pierluca.carenza@ba.infn.it}
\affiliation{Dipartimento Interateneo di Fisica ``Michelangelo Merlin,'' Via Amendola 173, 70126 Bari, Italy}
\affiliation{Istituto Nazionale di Fisica Nucleare - Sezione di Bari,
Via Orabona 4, 70126 Bari, Italy}             

\author{Bryce Fore}
\email{bryce4@uw.edu}
\affiliation{Institute for Nuclear Theory, University of Washington, Seattle, Washington 98195, USA}
\affiliation{Department of Physics, University of Washington, Seattle, Washington 98195, USA}

\author{Maurizio Giannotti}
\email{ mgiannotti@barry.edu}
\affiliation{Physical Sciences, Barry University, 11300 NE 2nd Avenue, Miami Shores, Florida 33161, USA}

\author{Alessandro Mirizzi}
\email{alessandro.mirizzi@ba.infn.it }
\affiliation{Dipartimento Interateneo di Fisica ``Michelangelo Merlin,'' Via Amendola 173, 70126 Bari, Italy}
\affiliation{Istituto Nazionale di Fisica Nucleare - Sezione di Bari,
Via Orabona 4, 70126 Bari, Italy}

\author{Sanjay Reddy}
\email{sareddy@uw.edu }
\affiliation{Institute for Nuclear Theory, University of Washington, Seattle, Washington 98195, USA}
\affiliation{Department of Physics, University of Washington, Seattle, Washington 98195, USA}


\begin{abstract}
We calculate the axion emission rate from reactions involving thermal pions in matter encountered in supernovae and neutron star mergers, identify unique spectral features, and explore their implications for astrophysics and particle physics. We find that it is about $2-5$ times larger than nucleon-nucleon bremsstrahlung, which in past studies was considered to be the dominant process. The axion spectrum is also found be much harder. Together, the larger rates and higher axion energies imply a stronger bound on the mass of the QCD axion and better prospects for direct detection in 
a large underground neutrino detector from a nearby galactic supernova.    

\end{abstract}

\maketitle

The axion, a hypothetical particle initially introduced to explain the smallness of the observed CP-violating interactions in QCD \cite{Weinberg:1977ma,Wilczek:1977pj} is a well-motivated dark matter (DM) candidate~\cite{Abbott:1982af,Preskill:1982cy,Dine:1982ah}. 
Axions produced during inflation would  account for the totality of the dark matter in the Universe if their mass is in the range from a few $\mu$eV to a few tens of  $\mu$eV~\cite{Borsanyi:2015cka,Ringwald:2016yge}, the exact value depending on unknown initial conditions. While this observation has motivated ongoing experimental searches for axions in the mass range $2~ \mu$eV$ \lesssim m_a \lesssim 25~ \mu$eV~\cite{Du:2018uak,Braine:2019fqb}, there is interest in axions with higher masses and experimental proposals to discover them~\cite{Brun:2019lyf,Irastorza:2018dyq,Sikivie:2020zpn} for two main reasons. 
First, recent work shows that if DM axions are produced after inflation, their mass needs to  be considerably larger to account for DM. When the contribution of topological defects to the axion production is properly accounted for in post-inflationary scenarios studies find that  $m_a \gtrsim 25~\mu\eV$ (see, e.g., \cite{DiVecchia:2019ejf} and references therein). Recent investigations suggest masses as high as $0.5-3.5 \meV$~\cite{Gorghetto:2018myk,Gorghetto:2020qws}, or even 15 meV~\cite{Ringwald:2015dsf}, depending on the specific axion model. Second, axion masses $m_a \gtrsim 1-10 \meV$ are particularly interesting for astrophysics, since these axions can have a noticeable impact on stellar evolution, supernovae, and the cooling of white dwarfs and neutron stars ~\cite{Raffelt:1996wa, Raffelt:2011ft,Giannotti:2015kwo,Giannotti:2017hny,DiLuzio:2020jjp,Isern:2018uce,Beznogov:2018fda}. 

The principle finding of this {Letter} is that the pion-induced axion emission from supernovae (SNe) provides new opportunities to either discover or constrain $\meV$ scale  axions.  We find that it strengthens the SN bound on axions and improves the prospect for both direct and indirect detection of SN axions in the parameter range of interest for particle physics, cosmology, and astrophysics.

The detection of about 20 neutrinos from the core-collapse SN  in the Large Magellanic Cloud in 1987, called SN 1987A, continues to provide one of the most stringent bounds  on the properties of the QCD axion. Pioneering work in Refs.~\cite{Raffelt:1987yt,Turner:1987by,Mayle:1987as} found that the axion emission due to nucleon-nucleon bremsstrahlung $NN \to NN a$ could dramatically alter the early cooling of a neutron star born with a fiducial temperature $T \simeq 30\MeV$ and change its neutrino luminosity. Subsequent improvements in the description of the axion emissivity from a SN core, over several years, demonstrated that the suppression of the neutrino luminosity due to axion emission would discernibly alter the observed neutrino events from SN 1987A to provide stringent bounds on the axion nucleon couplings \cite{Raffelt:1990yz,Engel:1990zd,Turner:1991ax,
Raffelt:1993ix,Keil:1996ju,Hanhart:2000ae,Fischer:2016cyd,Carenza:2019pxu}. This bound excludes QCD axions with masses in the range $15 \meV \lesssim m_a \lesssim 10 \keV $ \cite{Carenza:2019pxu}.

In all of these studies, the nucleon-nucleon bremsstrahlung reaction $NN \to NN a$ was assumed to be the dominant channel for the axion production in a SN core. The role of the pion-induced reaction,   $\pi^{-} p \to n a$ was first discussed in Refs.~\cite{Turner:1991ax,Raffelt:1993ix}, and in Ref.~\cite{Keil:1996ju} it was found to make the dominant contribution for a sufficiently high pion abundance. However, initial estimates suggested that the thermal pion population was too small for the pion reaction to be competitive  \cite{Raffelt:1996wa}. For this reason, pions and reactions involving pions in SNe have been largely ignored.   

A recent study demonstrated that the strong interactions enhance the abundance of negatively charged pions \cite{Fore:2019wib}. The study found that this enhancement can be reliably calculated for a wide range of density and temperature encountered in the SN core using the virial expansion. Motivated by this result, and by the large suppression of the bremsstrahlung rate found in~\cite{Carenza:2019pxu}, we revisit the calculation of the axion emissivity due to the reaction $\pi^{-} p \to n a$  to assess its impact. We find that, for pion densities predicted by the virial expansion, the pion-induced reaction dominates over the nucleon bremsstrahlung process over a wide range of ambient conditions and has important implications for the axion bounds derived from SN 1987A and direct detection in next-generation experiments. The enhanced emission also has implications for astrophysics of both core-collapse and neutron star mergers. In what follows, we describe our finding and these aforementioned implications.

To set the stage, we begin by briefly reviewing earlier work on axion production from nucleon bremsstrahlung reactions in SNe. First calculations of the bremsstrahlung rate were based on a simple model in which the nuclear interaction was described by the exchange of a virtual pion, often referred to as the one-pion-exchange (OPE) approximation \cite{Turner:1987by,Burrows:1988ah,Burrows:1990pk}. 
Furthermore, these studies neglected to properly account for the pion mass. 
In subsequent studies, a better treatment of the nuclear interaction beyond the OPE, which was consistent with nucleon-nucleon scattering data \cite{Hanhart:2000ae} and many-body corrections to the nucleon dispersion relations in the medium and its finite lifetime due to multiple scattering \cite{Raffelt:1991pw,Hannestad:1997gc,Raffelt:1996di}, was shown to reduce the axion emissivity.
The consistent inclusion of these effects led to an order of magnitude reduction in the axion emissivity relative to that obtained using the OPE prescription and implied a weaker bound on the axion mass~\cite{Carenza:2019pxu}. Reactions involving pions, as we demonstrate in this Letter, reverse this trend to strengthen the SN 1987A axion bound and improves the prospect for axion detection in large underground neutrino detectors.

Dense matter in the SN core is charge neutral, close to equilibrium with respect to weak interactions, and  characterized by a large isospin asymmetry. The difference between the neutron and proton chemical potentials, denoted by $\hat{\mu}=\mu_n-\mu_p$, increases with density and becomes comparable to  the pion mass $m_\pi \simeq 139$ MeV when the baryon density $n_B \gtrsim \nsat$, where $\nsat=1.6\times10^{38}\,{\rm cm}^{-3}$ is the saturation density (the corresponding mass density is $\rho_{\rm sat}\simeq 2.6 \times 10^{14} \gcm$). 
In the SN core, where neutrinos are trapped and weak equilibrium is quickly obtained, the pion chemical potential $\mu_{\pi^{-}}=\hat{\mu}=\mu_e-\mu_{\nu_e}$. When $\mu_\pi^{-}\simeq m_\pi$ the number density of negatively charged pions is greatly enhanced even when the ambient temperature realized in SNe, which is in the range of few MeV to few tens of MeV, is small compared to $m_\pi$. When $\mu_{\pi^{-}}> m_\pi$ a Bose-Einstein condensate of pions is favored, but whether or not this can be achieved at the densities encountered in SN matter is unclear \cite{Migdal:1990vm}. In what follows, we will only consider matter at densities where $\mu_{\pi^{-}}< m_\pi$. Under these conditions, the energy cost of introducing pions in dense matter is lowered by attractive p-wave interactions between nucleons and thermal pions (with typical momentum $p_\pi\simeq \sqrt{6 m_\pi T}\simeq 160{\rm MeV}~\sqrt{T/30~{\rm MeV}} $).

Although these mechanisms for enhancing the $\pi^-$ number density have been known for sometime, it is only recently that a model-independent calculation based on the virial expansion provided quantitative results when the $\pi^-$ fugacity denoted by $z_{\pi^-} = \exp{\left[\beta (\hat{\mu} - m_{\pi})\right]} \ll 1 $ \cite{Fore:2019wib}, 
where $\beta=1/T$, $T$ being the temperature. At leading order in the virial expansion, the number density of pions is given by 
\beq 
n_{\pi^-} =~z_{\pi} \left(  I_{\pi}  + \sum_{i=n,p} z_i ~b^{i\pi^-}_2+ {\cal O}(z^2_{i})\right) + {\cal O}(z^2_{\pi}) \,,	 
\label{eq:npi} 
\eeq  
where 
\beq 
I_{\pi}  = \int \frac{d^3k}{(2\pi)^3} \exp{\left[\beta( m_\pi-\sqrt{p^2+m^2_\pi} )\right]}\, 
\eeq 
is the contribution in the absence of interactions, and the second virial coefficients $ b^{n\pi^-}_2$ and $ b^{p\pi^-}_2$ include the contributions due to $\pi^-$ interactions with neutrons and protons, respectively.  It is adequate to retain only the leading term in the virial expansion when the fugacity of pions $z_{\pi^-}$ and neutrons and protons denoted by $z_n=\exp {\left[\beta(\mu_n-m_{n})\right]}$ and $z_p=\exp{\left[\beta(\mu_n-m_{p})\right]}$ are small compared to unity.   For a wide range of typical conditions encountered in a SN  where $z_{\pi^-} \ll 1$ and $z_p \ll1 $, and $z_n \lesssim 1 $, Eq.~(\ref{eq:npi}) provides a reliable estimate of the pion number density. For typical conditions encountered in the SN, the pion fraction $Y_\pi = n_{\pi^-}/n_B$, where $n_{\pi^-}$ is the pion number density and $n_{B}$ is the baryon density, was found to be in the range 1\%-5\% for $n_B\lesssim \nsat$. 

To describe reactions involving thermal pions, it is necessary to define the relation between the pion energy and its momentum given by 
\begin{align}
E_{\pi}(p)=\sqrt{p^{2}+m_{\pi}^{2}}+\Sigma(p)\,,
\label{eq:pi_dispersion} 
\end{align}
where $\Sigma(p)$ is the self-energy of pions at finite temperature and density, and incorporates the effects of pion interactions with nucleons. We employ a model in which the effective interaction between pions and nucleons is directly related to the measured pion-nucleon phase shifts (often called the pseudopotential) to calculate $\Sigma(p)$. The model is calibrated to reproduce the model-independent results obtained in the virial expansion and its use in calculating reactions is described in detail in Ref.~\cite{Fore:2019wib}.

The number of axions emitted per unit volume and per unit of time and energy is given by~\cite{Tamborra:2017ubu}
\begin{eqnarray}
\frac{d{\dot n}_a}{d\omega_{a}} &=& \int \frac{2d^{3}\bp_{p}}{(2\pi)^{3}2m_{N}}\frac{d^{3}\bp_{\pi}}{(2\pi)^{3}2E_{\pi}}\frac{2d^{3}\bp_{n}}{(2\pi)^{3}2m_{N}}\frac{4\pi \omega_{a}^{2}}{(2\pi)^{3}2\omega_{a}}
\nonumber \\
&\times &(2\pi)^{4}\delta^{4}(p_{f}-p_{i}) |\mathcal{{\overline M}}|^{2}f_{p}f_{\pi}(1-f_{n})\,\ . 
\label{eq:numbdens}
\end{eqnarray}
The squared transition matrix element 
in Eq.~(\ref{eq:numbdens})  is averaged over both initial and final nucleon spins and  given by  
\begin{align}
|\mathcal{{\overline M}}|^{2}= 4\bar{g}_{aN}^{2} \gamma_{\rm sf}(\omega_a) \left(\frac{g_{A}}{2F_{\pi}}\right)^{2}|\bp_{\pi}|^{2}\,,
\label{eq:matrix}
\end{align}
where  ${\bf p}_{\pi}$ is the pion momentum,
$g_A=1.26$ is the axial coupling, and $F_{\pi}= 92.4$ MeV is the pion decay constant. 
The effective axion-nucleon coupling $\bar{g}_{aN}$ is defined as
\begin{align}
\bar{g}_{aN}^{2} = g_a^{2}\left[\frac{1}{2}(C_{ap}^{2}+C_{an}^{2})+\frac{1}{3}C_{an}C_{ap}\right] \,,
\label{eq:effcoupl}
\end{align}
where $g_a= m_N/f_a$,  $m_N$ being the nucleon mass and $ f_a $ the Peccei-Quinn scale. We note  a discrepancy in Eq.~(\ref{eq:effcoupl}) with respect to the result~\cite{Turner:1991ax,Keil:1996ju}, i.e., a minus sign in front of the  $1/3 C_{an}C_{ap}$ term.
This difference arises because the mixed term in the matrix element is $-\frac{1}{2}C_{an}C_{ap}\left[2\langle(\hat{\bp}_{a}\cdot\hat{\bp}_{2})^{2}\rangle-1\right]$ and the average over the directions gives $\langle(\hat{\bp}_{a}\cdot\hat{\bp}_{2})^{2}\rangle=1/6$.
Depending on the axion couplings, this correction gives, at most, a difference of a factor 2 compared to previous literature.

The $C_{ai}$ are the model-dependent $\mathcal{O}(1)$ dimensionless axion-fermion couplings.
The couplings have been recently calculated for the KSVZ~\cite{Kim:1979if,Shifman:1979if} and the DFSZ~\cite{Zhitnitsky:1980tq,Dine:1981rt}  
models in Ref.~\cite{diCortona:2015ldu} (see~\cite{DiLuzio:2020wdo} for a discussion of these parameters in a large class of axion models). The function $\gamma_{\rm sf}(\omega_a)= \omega_a^2/[\omega_a^2+(\Gamma/2)^2]$
in Eq.~(\ref{eq:matrix}) is a simple ansatz suggested in Refs.~\cite{Raffelt:1993ix,Hannestad:1997gc} to account for the finite lifetime of the nucleon spin due to scattering in the dense medium, and $\Gamma$ is the nucleon spin fluctuation rate.  At a fiducial temperature $T=30 \MeV$ and mass density $\rho=10^{14} \gcm $, the calculations in \cite{Bartl:2014hoa,Carenza:2019pxu} indicate that $\Gamma \simeq 35 \MeV$.

The distribution functions of the different interacting species are the usual Fermi-Dirac or Bose-Einstein distribution,
\begin{align}
f_i(E)=\frac{1}{e^{\left[ E_i(p_i)-\mu_i\right] /T} \mp1}\;,
\end{align}
where the $+$ sign applies to fermions, while the $-$  is for bosons, and $\mu_i$ are the chemical potentials for $i=p,n,\pi$. Corrections to the dispersion relations $E_i(p_i)$ of nucleons are incorporated through the equation  
 \beq
 E_i = m_N + \frac{{|{\bf p}_i|}^{2}}{2 m^{\ast}_N} +U_i \,\ ,
 \label{eq:energy}
 \eeq
where the nucleon effective mass $m^{\ast}_N$ and single-particle potentials $U_i$ are obtained from Ref.~\cite{Fore:2019wib}. The modification to the pion dispersion relation due to its interactions with nucleons is incorporated through Eq.~(\ref{eq:pi_dispersion}) 
with $\Sigma(p)$ obtained consistently as described in Ref.~\cite{Fore:2019wib}. 

The differential axion number luminosity, which is defined to be the total number of axions emitted in a specified energy range per unit time from the SN is obtained by integrating Eq.~(\ref{eq:numbdens}) over
the SN volume and is given by
\begin{equation}
\frac{d{\cal N}_a}{d\omega_{a}} =\int d^{3}r\,\frac{d{\dot n}_a}{d\omega_{a}}\;.    
\end{equation}

The energy radiated in axions per unit volume and time, called the axion emissivity, can be calculated directly from Eq.~(\ref{eq:numbdens}) as
\begin{equation}
Q_{a}=\int d\omega_{a}\omega_{a}\frac{d{\dot n}_a}{d\omega_{a}}\,,
\label{eq:emiss}
\end{equation}
where the phase-space integrals can be performed to obtain
a simpler expression for pionic processes
\begin{eqnarray}
Q^{\pi}_{a}&=&
\frac{\bar{g}_{aN}^{2}T^{7.5}}{\sqrt{2\,m_N}\,\pi^{5}}
\left(\frac{g_{A}}{2F_{\pi}}\right)^{2}\frac{z_\pi\, z_p}{1+z_n}
\left[\int dx_{p}\frac{ x_{p}^{2}}{e^{x_{p}^{2}}+z_p}\right] \nonumber \\
& &\int dx_{\pi}\frac{x_{\pi}^{3}\epsilon_{\pi}^{2}}{\left( e^{\epsilon_\pi-y_\pi}-z_\pi \right)} \frac{\epsilon_\pi^{2}}{ \left[ \epsilon_\pi^{2}+(\Gamma/2T)^{2}\right] } \,,
\label{eq:emisspion}
\end{eqnarray}
with $x_p=|\bp_{p}|/\sqrt{2m_{N}T}$, $x_{\pi}=|\bp_{\pi}|/T$, $y_{\pi}=m_{\pi}/T$, and $\epsilon_\pi=E_\pi/T $. The fugacities $z_\pi$ and $z_p$ were defined earlier. 
Finally,  the total axion energy luminosity is given by
\begin{equation}
L_a = \int d^{3}r\, Q_a(r) \,\ .  
\end{equation}

The enhancement of the axion emission rate due to the pion reaction relative to the bremsstrahlung calculated 
in~\cite{Carenza:2019pxu} can be gauged from Table \ref{tab:emissiv}, where we compare the $\pi N$ and $NN$ axion emissivity at different post-bounce times using ambient conditions taken from the SN model described in~\cite{Carenza:2019pxu} at a specific radial location $r=10\,{\rm km}$. We estimate the total axion emissivity $L_a$ by assuming average values for $T$ and $\rho$ within the region $r< 12$ km. This is shown in the last column of the Table.
We realize that the axion emissivity is increased by factor of about 4 due to pionic reactions at $t_{\rm pb}=1$ s. At later times, the pion contribution is less important: the total emissivity is only a factor 2 larger
than the one from NN process for $t_{\rm pb}=6$ s.
\begin{table*}
 \caption{Axion emissivities $Q_{a}$ in units of $10^{32} \erg \cm^{-3}\s^{-1}$ and luminosities $L_a$ in units $10^{51} \erg\s^{-1}$
 for KSVZ model ($C_{ap}=-0.47\; ; C_{an}=0$)  and $g_{a}=m_{N}/f_{a}=10^{-9}$, for different post-bounce times.
 }
\begin{tabular}{lccccccc}
\hline
$t_{\rm pb} \,\  $  &$\rho \,\  $&T&$Y_{\pi}$ & $Q_{a}^{NN} $&$Q_{a}^{\pi}$&$Q_{a}^{\rm tot}/Q_{a}^{NN}$&
$L_a$\\
(s)  & $(10^{14} \textrm{g cm}^{-3}) \,\ $& (MeV) & & $( 10^{32} \erg \cm^{-3}\s^{-1}) \,\ $ & $(10^{32} \erg \cm^{-3}\s^{-1})$ & &$( 10^{51} \erg\s^{-1})$ \\
\hline
1&1.45&37.07  \,\ &0.011&1.37&4.63&4.38&4.0\\
2&2.08&38.93  \,\ &0.016&3.28&8.87&3.70&8.10\\
4&3.10&40.56  \,\ &0.027&9.08&15.87&2.75&16.63\\
6&3.65&39.91  \,\ &0.034&12.92&14.99&2.16&18.61\\
\hline
\end{tabular}
\label{tab:emissiv}
\end{table*}
\begin{table*}[!t]
 \caption{
 Bound on the effective axion-nucleon coupling $\bar{g}_{aN}$ obtained using Eq.~(\ref{eq:Raffelt}). The corresponding bound on $m_a$ and $f_a$ for KSVZ model with $C_{ap}=-0.47\,, C_{an}=0$ are also shown.}
\begin{center}
\begin{tabular}{lcccc}
\hline
$\rho$ & &
$\,\ \bar{g}_{aN}$
& $m_{a}$ &$f_{a}$\\
 & & ($\times10^{-9}$) & \,\ (meV) & \,\ $ (\times10^{8}$~GeV) \\
\hline
\hline
$\rho_{0}$&only $NN$ &0.81  &21.02 &2.71 \\
&$\pi N+NN$&0.46   &11.99 &4.75 \\
$\rho_{0}/2$&only $NN$ & 0.93&24.11   &2.36   \\
&$\pi N+NN$&0.42  &10.96    &5.20    \\
\hline
\end{tabular}
\label{tab:massbound}
\end{center}
\end{table*}

The more stringent bound on the axion mass implied by the larger emissivity can be estimated using an observation made by Raffelt \cite{Raffelt:2006cw} who found that, for 
\begin{align}
\frac{Q_{a}}{\rho}>10^{19}\erg\g^{-1}\s^{-1}\;,
\label{eq:Raffelt}
\end{align}
simulations predicted a significant shortening of the SN 1987A neutrino signal. The axion emissivity is typically calculated at a fiducial density  $\rho=\rho_{\rm sat}$,  $T=30\MeV$, and proton fraction $Y_{p}=0.3$.  In Table~\ref{tab:massbound} we show the bounds derived for the KSVZ axion  obtained using the fiducial densities $\rho=\rho_{\rm sat}$
and $\rho=\rho_{\rm sat}/2$ at temperature $T=30\MeV$ and proton fraction $Y_{p}=0.3$. Since the rates are $\propto m_a^2$, the factor of $4$ enhancement in the rate strengthens the axion mass bound by a  factor $2$. 

In the DFSZ model, the axion-nucleon couplings are expressed as a function of  $\tan\beta\equiv v_u/v_d$, which
represents the ratio of the two Higgs bosons in the model and is  constrained in the range $0.25<\tan\beta<170$ \cite{DiLuzio:2020wdo}. Correspondingly,  in this case for $\rho=\rho_{\rm sat}$ when including pions, the axion mass bound is shifted  from $9.3\meV<m_{a}<17.7\meV$ to $5.8\meV<m_{a}<10.9\meV$.

We caution the reader that, while this simple estimate captures that trend and the relative importance of the pion reaction, detailed SN simulations with pions will be needed to derive a robust bound.

We also remark that 
in the mass range of interest axions are not trapped in the SN core. This can be shown by calculating the mean free path for this process. Following \cite{Raffelt:1996wa}, we obtain 
\begin{equation}
l_{\pi}^{-1}=\frac{\bar{g}_{aN}^{2}\pi}{4m_{N}^{4}}\left(\frac{g_{A}}{2F_{\pi}}\right)^{2}\frac{\rho^{2}Y_{p}Y_{\pi}}{T}
\end{equation}
in the non-degenerate limit for nucleons and pions, neglecting the pion mass and the multiple nucleon scattering effect. The first approximation is reasonable in the region around the SN core; the other two approximations are conservative since that would  only reduce the mean free path. For typical SN conditions, the mean free path results $l_{\pi}\sim (\bar{g}_{aN}/10^{-9})^{-2}\,10^{5}\km$. This finding confirms that  axions are in the free-streaming regime for $\bar{g}_{aN}<10^{-7}$.

\begin{figure}[!t]
\vspace{0.cm}
\includegraphics[width=0.8\columnwidth]{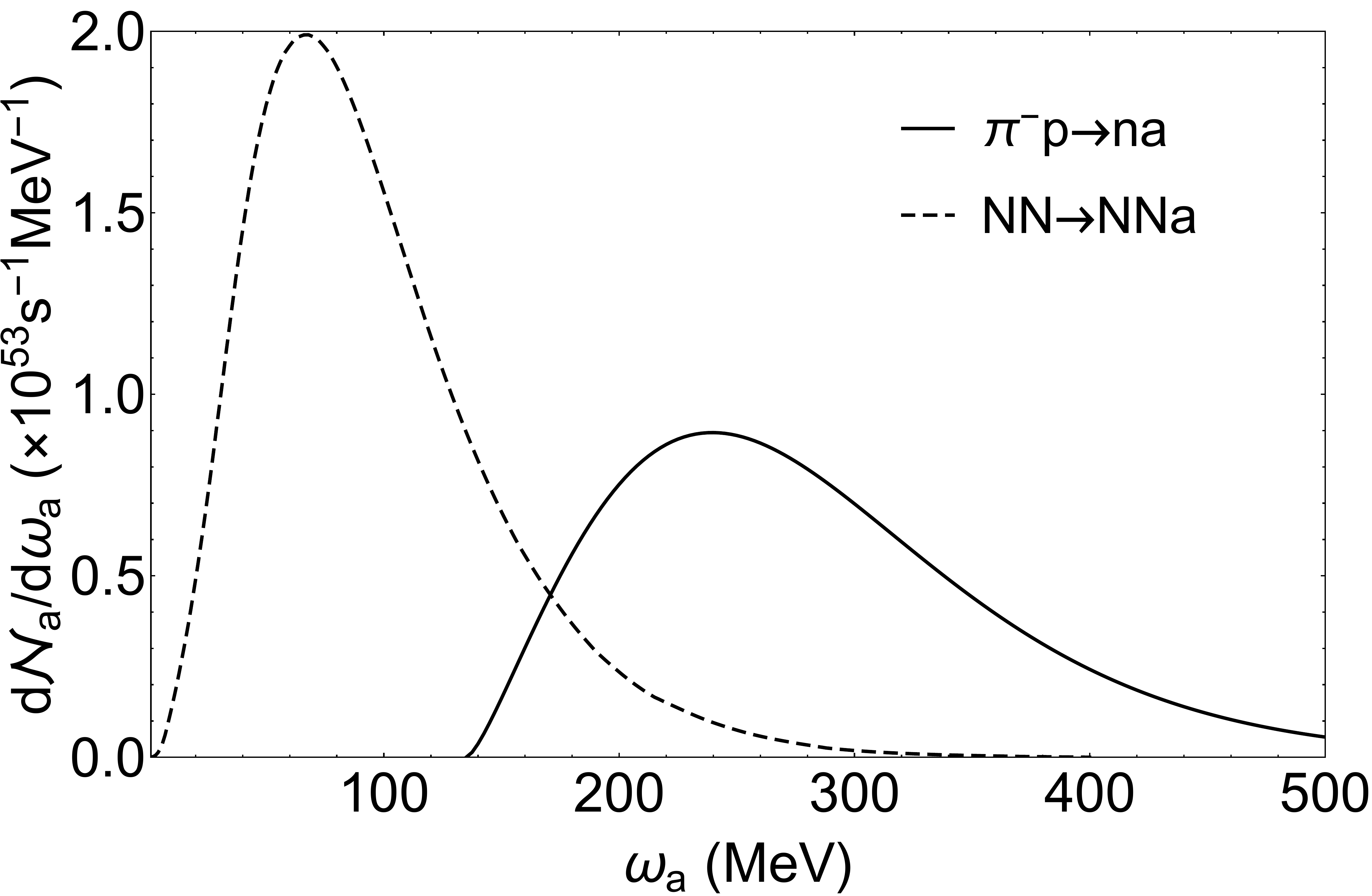}
\caption{The number spectra of axions  for $\pi N$ (solid curve) and $NN$ (dashed  curve) processes for our benchmark axion model at a post-bounce time $t_{\rm pb}=1\s$.}
\label{fig:pionflux}
\end{figure}

In addition to increasing the total axion emissivity, the reaction involving pions produces axions with a harder energy spectrum. This is to be expected as these reactions harness the rest mass energy of the pion in the initial state. Fig.~\ref{fig:pionflux} compares the axion number luminosity obtained from pionic reactions (solid curve) to those from nucleon bremsstrahlung (dashed curve) for our benchmark axion model at a post-bounce time $t_{\rm pb}=1\s$. 

The larger axion energies, especially axions in the range $200-300 \MeV$ are particularly interesting for detection in neutrino underground experiments. This is because at these energies we expect a resonant enhancement of the axion-nucleon cross section due to the $\Delta$ intermediate state. These high energy axions can produce neutral and charged pions in water Cherenkov detectors due to the reactions $a+p \rightarrow p + \pi^0$, $a+p \rightarrow n + \pi^+$, and $a+n \rightarrow p + \pi^-$. The operator structure that describes axion coupling to nucleons is nearly identical to the pion-nucleon coupling, but with $f_\pi$ replaced by $f_a$. This observation has been used earlier to suggest that the cross section for the reaction $a + p \rightarrow N + \pi$, 
$\sigma_{aN} \simeq (F_\pi/f_a)^2 \sigma_{\pi N}$ where $\sigma_{\pi N}$ is the cross section for $\pi^0 + p \rightarrow p + \pi^0$ \cite{Raffelt:2011ft}. In the resonance region, which can be accessed when the axion energy $E_a\simeq 200-300 \MeV$, the cross-section $\sigma_{\pi N} \approx 100$ mb. For $f_a=10^9\GeV$ ($m_a=5.7 \meV$), an order of magnitude estimate obtained using the axion luminosity in Table \ref{tab:emissiv} suggests that about $1000$ pions will be produced in a megaton water Cherenkov detector for a SN at $1 \kpc$.  

This intriguing prospect for direct detection of axions from a galactic SN warrants further studies and our Letter identifies several directions for future research. Most importantly, it motivates rigorous calculations of the cross section for the process $a + p \rightarrow N + \pi$, as this is critical for the pion production in water Cherenkov detectors. Such calculations will also address possible resonant enhancement of the inverse reaction $\pi^-+ p \rightarrow n + a$ in the SN environment. Further work, which goes beyond the virial expansion in Ref.~\cite{Fore:2019wib}, is needed to assess how the pion abundances increase with density in the SN core. Although our initial estimates suggest an exponential increase of the pion thermal population with density, reliable calculations that can accommodate Bose-Einstein condensation of pions at finite temperature will be needed in this context (for a discussion of meson condensation in SN matter, see Ref.~\cite{Pons:2000iy}). Ultimately, advanced SN simulations that incorporate the pion contribution to both thermodynamics and reactions will be essential to fully assess the impact of the enhanced axion luminosity and energies that we discuss in this Letter.

\textit{Acknowledgments---} 
A.M. warmly thanks John Beacom and Mark Vagins for useful discussions on the possibility of SN axion detection in a large underground neutrino detector. We also thank Georg 
Raffelt for valuable comments 
on the Letter.
The work of P.C. and 
A.M. is partially supported by the Italian Istituto Nazionale di Fisica Nucleare (INFN) through the ``Theoretical Astroparticle Physics'' project
and by the research grant number 2017W4HA7S
``NAT-NET: Neutrino and Astroparticle Theory Network'' under the program
PRIN 2017 funded by the Italian Ministero dell'Universit\`a e della
Ricerca (MUR). S. R. was supported by the Department of Energy grant DE-FG02-00ER41132. B. F. acknowledges support from the SciDAC grant A18-0354-S002 (de-sc0018232) and by the National Science Foundation Graduate Research Fellowship Program grant DGE-1762114.


\bibliography{references}

\begin{thebibliography}{51}
\expandafter\ifx\csname natexlab\endcsname\relax\def\natexlab#1{#1}\fi
\expandafter\ifx\csname bibnamefont\endcsname\relax
  \def\bibnamefont#1{#1}\fi
\expandafter\ifx\csname bibfnamefont\endcsname\relax
  \def\bibfnamefont#1{#1}\fi
\expandafter\ifx\csname citenamefont\endcsname\relax
  \def\citenamefont#1{#1}\fi
\expandafter\ifx\csname url\endcsname\relax
  \def\url#1{\texttt{#1}}\fi
\expandafter\ifx\csname urlprefix\endcsname\relax\def\urlprefix{URL }\fi
\providecommand{\bibinfo}[2]{#2}
\providecommand{\eprint}[2][]{\url{#2}}

\bibitem[{\citenamefont{Weinberg}(1978)}]{Weinberg:1977ma}
\bibinfo{author}{\bibfnamefont{S.}~\bibnamefont{Weinberg}},
  \bibinfo{journal}{Phys. Rev. Lett.} \textbf{\bibinfo{volume}{40}},
  \bibinfo{pages}{223} (\bibinfo{year}{1978}).

\bibitem[{\citenamefont{Wilczek}(1978)}]{Wilczek:1977pj}
\bibinfo{author}{\bibfnamefont{F.}~\bibnamefont{Wilczek}},
  \bibinfo{journal}{Phys. Rev. Lett.} \textbf{\bibinfo{volume}{40}},
  \bibinfo{pages}{279} (\bibinfo{year}{1978}).

\bibitem[{\citenamefont{Abbott and Sikivie}(1983)}]{Abbott:1982af}
\bibinfo{author}{\bibfnamefont{L.}~\bibnamefont{Abbott}} \bibnamefont{and}
  \bibinfo{author}{\bibfnamefont{P.}~\bibnamefont{Sikivie}},
  \bibinfo{journal}{Phys. Lett. B} \textbf{\bibinfo{volume}{120}},
  \bibinfo{pages}{133} (\bibinfo{year}{1983}).

\bibitem[{\citenamefont{Preskill et~al.}(1983)\citenamefont{Preskill, Wise, and
  Wilczek}}]{Preskill:1982cy}
\bibinfo{author}{\bibfnamefont{J.}~\bibnamefont{Preskill}},
  \bibinfo{author}{\bibfnamefont{M.~B.} \bibnamefont{Wise}}, \bibnamefont{and}
  \bibinfo{author}{\bibfnamefont{F.}~\bibnamefont{Wilczek}},
  \bibinfo{journal}{Phys. Lett. B} \textbf{\bibinfo{volume}{120}},
  \bibinfo{pages}{127} (\bibinfo{year}{1983}).

\bibitem[{\citenamefont{Dine and Fischler}(1983)}]{Dine:1982ah}
\bibinfo{author}{\bibfnamefont{M.}~\bibnamefont{Dine}} \bibnamefont{and}
  \bibinfo{author}{\bibfnamefont{W.}~\bibnamefont{Fischler}},
  \bibinfo{journal}{Phys. Lett. B} \textbf{\bibinfo{volume}{120}},
  \bibinfo{pages}{137} (\bibinfo{year}{1983}).

\bibitem[{\citenamefont{Borsanyi et~al.}(2016)\citenamefont{Borsanyi, Dierigl,
  Fodor, Katz, Mages, Nogradi, Redondo, Ringwald, and
  Szabo}}]{Borsanyi:2015cka}
\bibinfo{author}{\bibfnamefont{S.}~\bibnamefont{Borsanyi}},
  \bibinfo{author}{\bibfnamefont{M.}~\bibnamefont{Dierigl}},
  \bibinfo{author}{\bibfnamefont{Z.}~\bibnamefont{Fodor}},
  \bibinfo{author}{\bibfnamefont{S.}~\bibnamefont{Katz}},
  \bibinfo{author}{\bibfnamefont{S.}~\bibnamefont{Mages}},
  \bibinfo{author}{\bibfnamefont{D.}~\bibnamefont{Nogradi}},
  \bibinfo{author}{\bibfnamefont{J.}~\bibnamefont{Redondo}},
  \bibinfo{author}{\bibfnamefont{A.}~\bibnamefont{Ringwald}}, \bibnamefont{and}
  \bibinfo{author}{\bibfnamefont{K.}~\bibnamefont{Szabo}},
  \bibinfo{journal}{Phys. Lett. B} \textbf{\bibinfo{volume}{752}},
  \bibinfo{pages}{175} (\bibinfo{year}{2016}), \eprint{1508.06917}.

\bibitem[{\citenamefont{Ringwald}(2016)}]{Ringwald:2016yge}
\bibinfo{author}{\bibfnamefont{A.}~\bibnamefont{Ringwald}},
  \bibinfo{journal}{PoS} \textbf{\bibinfo{volume}{NOW2016}},
  \bibinfo{pages}{081} (\bibinfo{year}{2016}), \eprint{1612.08933}.

\bibitem[{\citenamefont{Du et~al.}(2018)}]{Du:2018uak}
\bibinfo{author}{\bibfnamefont{N.}~\bibnamefont{Du}} \bibnamefont{et~al.}
  (\bibinfo{collaboration}{ADMX}), \bibinfo{journal}{Phys. Rev. Lett.}
  \textbf{\bibinfo{volume}{120}}, \bibinfo{pages}{151301}
  (\bibinfo{year}{2018}), \eprint{1804.05750}.

\bibitem[{\citenamefont{Braine et~al.}(2020)}]{Braine:2019fqb}
\bibinfo{author}{\bibfnamefont{T.}~\bibnamefont{Braine}} \bibnamefont{et~al.}
  (\bibinfo{collaboration}{ADMX}), \bibinfo{journal}{Phys. Rev. Lett.}
  \textbf{\bibinfo{volume}{124}}, \bibinfo{pages}{101303}
  (\bibinfo{year}{2020}), \eprint{1910.08638}.

\bibitem[{\citenamefont{Brun et~al.}(2019)}]{Brun:2019lyf}
\bibinfo{author}{\bibfnamefont{P.}~\bibnamefont{Brun}} \bibnamefont{et~al.}
  (\bibinfo{collaboration}{MADMAX}), \bibinfo{journal}{Eur. Phys. J. C}
  \textbf{\bibinfo{volume}{79}}, \bibinfo{pages}{186} (\bibinfo{year}{2019}),
  \eprint{1901.07401}.

\bibitem[{\citenamefont{Irastorza and Redondo}(2018)}]{Irastorza:2018dyq}
\bibinfo{author}{\bibfnamefont{I.~G.} \bibnamefont{Irastorza}}
  \bibnamefont{and} \bibinfo{author}{\bibfnamefont{J.}~\bibnamefont{Redondo}},
  \bibinfo{journal}{Prog. Part. Nucl. Phys.} \textbf{\bibinfo{volume}{102}},
  \bibinfo{pages}{89} (\bibinfo{year}{2018}), \eprint{1801.08127}.

\bibitem[{\citenamefont{Sikivie}(2020)}]{Sikivie:2020zpn}
\bibinfo{author}{\bibfnamefont{P.}~\bibnamefont{Sikivie}}
  (\bibinfo{year}{2020}), \eprint{2003.02206}.

\bibitem[{\citenamefont{Di~Vecchia et~al.}(2019)\citenamefont{Di~Vecchia,
  Giannotti, Lattanzi, and Lindner}}]{DiVecchia:2019ejf}
\bibinfo{author}{\bibfnamefont{P.}~\bibnamefont{Di~Vecchia}},
  \bibinfo{author}{\bibfnamefont{M.}~\bibnamefont{Giannotti}},
  \bibinfo{author}{\bibfnamefont{M.}~\bibnamefont{Lattanzi}}, \bibnamefont{and}
  \bibinfo{author}{\bibfnamefont{A.}~\bibnamefont{Lindner}},
  \bibinfo{journal}{PoS} \textbf{\bibinfo{volume}{Confinement2018}},
  \bibinfo{pages}{034} (\bibinfo{year}{2019}), \eprint{1902.06567}.

\bibitem[{\citenamefont{Gorghetto et~al.}(2018)\citenamefont{Gorghetto, Hardy,
  and Villadoro}}]{Gorghetto:2018myk}
\bibinfo{author}{\bibfnamefont{M.}~\bibnamefont{Gorghetto}},
  \bibinfo{author}{\bibfnamefont{E.}~\bibnamefont{Hardy}}, \bibnamefont{and}
  \bibinfo{author}{\bibfnamefont{G.}~\bibnamefont{Villadoro}},
  \bibinfo{journal}{JHEP} \textbf{\bibinfo{volume}{07}}, \bibinfo{pages}{151}
  (\bibinfo{year}{2018}), \eprint{1806.04677}.

\bibitem[{\citenamefont{Gorghetto et~al.}(2020)\citenamefont{Gorghetto, Hardy,
  and Villadoro}}]{Gorghetto:2020qws}
\bibinfo{author}{\bibfnamefont{M.}~\bibnamefont{Gorghetto}},
  \bibinfo{author}{\bibfnamefont{E.}~\bibnamefont{Hardy}}, \bibnamefont{and}
  \bibinfo{author}{\bibfnamefont{G.}~\bibnamefont{Villadoro}}
  (\bibinfo{year}{2020}), \eprint{2007.04990}.

\bibitem[{\citenamefont{Ringwald and Saikawa}(2016)}]{Ringwald:2015dsf}
\bibinfo{author}{\bibfnamefont{A.}~\bibnamefont{Ringwald}} \bibnamefont{and}
  \bibinfo{author}{\bibfnamefont{K.}~\bibnamefont{Saikawa}},
  \bibinfo{journal}{Phys. Rev. D} \textbf{\bibinfo{volume}{93}},
  \bibinfo{pages}{085031} (\bibinfo{year}{2016}), \bibinfo{note}{[Addendum:
  Phys.Rev.D 94, 049908 (2016)]}, \eprint{1512.06436}.

\bibitem[{\citenamefont{Raffelt}(1996)}]{Raffelt:1996wa}
\bibinfo{author}{\bibfnamefont{G.}~\bibnamefont{Raffelt}},
  \emph{\bibinfo{title}{{Stars as laboratories for fundamental physics}}}
  (\bibinfo{year}{1996}), ISBN \bibinfo{isbn}{978-0-226-70272-8}.

\bibitem[{\citenamefont{Raffelt et~al.}(2011)\citenamefont{Raffelt, Redondo,
  and Viaux~Maira}}]{Raffelt:2011ft}
\bibinfo{author}{\bibfnamefont{G.~G.} \bibnamefont{Raffelt}},
  \bibinfo{author}{\bibfnamefont{J.}~\bibnamefont{Redondo}}, \bibnamefont{and}
  \bibinfo{author}{\bibfnamefont{N.}~\bibnamefont{Viaux~Maira}},
  \bibinfo{journal}{Phys. Rev. D} \textbf{\bibinfo{volume}{84}},
  \bibinfo{pages}{103008} (\bibinfo{year}{2011}), \eprint{1110.6397}.

\bibitem[{\citenamefont{Giannotti et~al.}(2016)\citenamefont{Giannotti,
  Irastorza, Redondo, and Ringwald}}]{Giannotti:2015kwo}
\bibinfo{author}{\bibfnamefont{M.}~\bibnamefont{Giannotti}},
  \bibinfo{author}{\bibfnamefont{I.}~\bibnamefont{Irastorza}},
  \bibinfo{author}{\bibfnamefont{J.}~\bibnamefont{Redondo}}, \bibnamefont{and}
  \bibinfo{author}{\bibfnamefont{A.}~\bibnamefont{Ringwald}},
  \bibinfo{journal}{JCAP} \textbf{\bibinfo{volume}{05}}, \bibinfo{pages}{057}
  (\bibinfo{year}{2016}), \eprint{1512.08108}.

\bibitem[{\citenamefont{Giannotti et~al.}(2017)\citenamefont{Giannotti,
  Irastorza, Redondo, Ringwald, and Saikawa}}]{Giannotti:2017hny}
\bibinfo{author}{\bibfnamefont{M.}~\bibnamefont{Giannotti}},
  \bibinfo{author}{\bibfnamefont{I.~G.} \bibnamefont{Irastorza}},
  \bibinfo{author}{\bibfnamefont{J.}~\bibnamefont{Redondo}},
  \bibinfo{author}{\bibfnamefont{A.}~\bibnamefont{Ringwald}}, \bibnamefont{and}
  \bibinfo{author}{\bibfnamefont{K.}~\bibnamefont{Saikawa}},
  \bibinfo{journal}{JCAP} \textbf{\bibinfo{volume}{10}}, \bibinfo{pages}{010}
  (\bibinfo{year}{2017}), \eprint{1708.02111}.

\bibitem[{\citenamefont{Di~Luzio
  et~al.}(2020{\natexlab{a}})\citenamefont{Di~Luzio, Fedele, Giannotti, Mescia,
  and Nardi}}]{DiLuzio:2020jjp}
\bibinfo{author}{\bibfnamefont{L.}~\bibnamefont{Di~Luzio}},
  \bibinfo{author}{\bibfnamefont{M.}~\bibnamefont{Fedele}},
  \bibinfo{author}{\bibfnamefont{M.}~\bibnamefont{Giannotti}},
  \bibinfo{author}{\bibfnamefont{F.}~\bibnamefont{Mescia}}, \bibnamefont{and}
  \bibinfo{author}{\bibfnamefont{E.}~\bibnamefont{Nardi}}
  (\bibinfo{year}{2020}{\natexlab{a}}), \eprint{2006.12487}.

\bibitem[{\citenamefont{Isern et~al.}(2018)\citenamefont{Isern, Garcia-Berro,
  Torres, Cojocaru, and Catalan}}]{Isern:2018uce}
\bibinfo{author}{\bibfnamefont{J.}~\bibnamefont{Isern}},
  \bibinfo{author}{\bibfnamefont{E.}~\bibnamefont{Garcia-Berro}},
  \bibinfo{author}{\bibfnamefont{S.}~\bibnamefont{Torres}},
  \bibinfo{author}{\bibfnamefont{R.}~\bibnamefont{Cojocaru}}, \bibnamefont{and}
  \bibinfo{author}{\bibfnamefont{S.}~\bibnamefont{Catalan}},
  \bibinfo{journal}{Mon. Not. Roy. Astron. Soc.}
  \textbf{\bibinfo{volume}{478}}, \bibinfo{pages}{2569} (\bibinfo{year}{2018}),
  \eprint{1805.00135}.

\bibitem[{\citenamefont{Beznogov et~al.}(2018)\citenamefont{Beznogov, Rrapaj,
  Page, and Reddy}}]{Beznogov:2018fda}
\bibinfo{author}{\bibfnamefont{M.~V.} \bibnamefont{Beznogov}},
  \bibinfo{author}{\bibfnamefont{E.}~\bibnamefont{Rrapaj}},
  \bibinfo{author}{\bibfnamefont{D.}~\bibnamefont{Page}}, \bibnamefont{and}
  \bibinfo{author}{\bibfnamefont{S.}~\bibnamefont{Reddy}},
  \bibinfo{journal}{Phys. Rev. C} \textbf{\bibinfo{volume}{98}},
  \bibinfo{pages}{035802} (\bibinfo{year}{2018}), \eprint{1806.07991}.

\bibitem[{\citenamefont{Raffelt and Seckel}(1988)}]{Raffelt:1987yt}
\bibinfo{author}{\bibfnamefont{G.}~\bibnamefont{Raffelt}} \bibnamefont{and}
  \bibinfo{author}{\bibfnamefont{D.}~\bibnamefont{Seckel}},
  \bibinfo{journal}{Phys. Rev. Lett.} \textbf{\bibinfo{volume}{60}},
  \bibinfo{pages}{1793} (\bibinfo{year}{1988}).

\bibitem[{\citenamefont{Turner}(1988)}]{Turner:1987by}
\bibinfo{author}{\bibfnamefont{M.~S.} \bibnamefont{Turner}},
  \bibinfo{journal}{Phys. Rev. Lett.} \textbf{\bibinfo{volume}{60}},
  \bibinfo{pages}{1797} (\bibinfo{year}{1988}).

\bibitem[{\citenamefont{Mayle et~al.}(1988)\citenamefont{Mayle, Wilson, Ellis,
  Olive, Schramm, and Steigman}}]{Mayle:1987as}
\bibinfo{author}{\bibfnamefont{R.}~\bibnamefont{Mayle}},
  \bibinfo{author}{\bibfnamefont{J.~R.} \bibnamefont{Wilson}},
  \bibinfo{author}{\bibfnamefont{J.~R.} \bibnamefont{Ellis}},
  \bibinfo{author}{\bibfnamefont{K.~A.} \bibnamefont{Olive}},
  \bibinfo{author}{\bibfnamefont{D.~N.} \bibnamefont{Schramm}},
  \bibnamefont{and} \bibinfo{author}{\bibfnamefont{G.}~\bibnamefont{Steigman}},
  \bibinfo{journal}{Phys. Lett. B} \textbf{\bibinfo{volume}{203}},
  \bibinfo{pages}{188} (\bibinfo{year}{1988}).

\bibitem[{\citenamefont{Raffelt}(1990)}]{Raffelt:1990yz}
\bibinfo{author}{\bibfnamefont{G.~G.} \bibnamefont{Raffelt}},
  \bibinfo{journal}{Phys. Rept.} \textbf{\bibinfo{volume}{198}},
  \bibinfo{pages}{1} (\bibinfo{year}{1990}).

\bibitem[{\citenamefont{Engel et~al.}(1990)\citenamefont{Engel, Seckel, and
  Hayes}}]{Engel:1990zd}
\bibinfo{author}{\bibfnamefont{J.}~\bibnamefont{Engel}},
  \bibinfo{author}{\bibfnamefont{D.}~\bibnamefont{Seckel}}, \bibnamefont{and}
  \bibinfo{author}{\bibfnamefont{A.}~\bibnamefont{Hayes}},
  \bibinfo{journal}{Phys. Rev. Lett.} \textbf{\bibinfo{volume}{65}},
  \bibinfo{pages}{960} (\bibinfo{year}{1990}).

\bibitem[{\citenamefont{Turner}(1992)}]{Turner:1991ax}
\bibinfo{author}{\bibfnamefont{M.~S.} \bibnamefont{Turner}},
  \bibinfo{journal}{Phys. Rev. D} \textbf{\bibinfo{volume}{45}},
  \bibinfo{pages}{1066} (\bibinfo{year}{1992}).

\bibitem[{\citenamefont{Raffelt and Seckel}(1995)}]{Raffelt:1993ix}
\bibinfo{author}{\bibfnamefont{G.}~\bibnamefont{Raffelt}} \bibnamefont{and}
  \bibinfo{author}{\bibfnamefont{D.}~\bibnamefont{Seckel}},
  \bibinfo{journal}{Phys. Rev. D} \textbf{\bibinfo{volume}{52}},
  \bibinfo{pages}{1780} (\bibinfo{year}{1995}), \eprint{astro-ph/9312019}.

\bibitem[{\citenamefont{Keil et~al.}(1997)\citenamefont{Keil, Janka, Schramm,
  Sigl, Turner, and Ellis}}]{Keil:1996ju}
\bibinfo{author}{\bibfnamefont{W.}~\bibnamefont{Keil}},
  \bibinfo{author}{\bibfnamefont{H.-T.} \bibnamefont{Janka}},
  \bibinfo{author}{\bibfnamefont{D.~N.} \bibnamefont{Schramm}},
  \bibinfo{author}{\bibfnamefont{G.}~\bibnamefont{Sigl}},
  \bibinfo{author}{\bibfnamefont{M.~S.} \bibnamefont{Turner}},
  \bibnamefont{and} \bibinfo{author}{\bibfnamefont{J.~R.} \bibnamefont{Ellis}},
  \bibinfo{journal}{Phys. Rev. D} \textbf{\bibinfo{volume}{56}},
  \bibinfo{pages}{2419} (\bibinfo{year}{1997}), \eprint{astro-ph/9612222}.

\bibitem[{\citenamefont{Hanhart et~al.}(2001)\citenamefont{Hanhart, Phillips,
  and Reddy}}]{Hanhart:2000ae}
\bibinfo{author}{\bibfnamefont{C.}~\bibnamefont{Hanhart}},
  \bibinfo{author}{\bibfnamefont{D.~R.} \bibnamefont{Phillips}},
  \bibnamefont{and} \bibinfo{author}{\bibfnamefont{S.}~\bibnamefont{Reddy}},
  \bibinfo{journal}{Phys. Lett. B} \textbf{\bibinfo{volume}{499}},
  \bibinfo{pages}{9} (\bibinfo{year}{2001}), \eprint{astro-ph/0003445}.

\bibitem[{\citenamefont{Fischer et~al.}(2016)\citenamefont{Fischer,
  Chakraborty, Giannotti, Mirizzi, Payez, and Ringwald}}]{Fischer:2016cyd}
\bibinfo{author}{\bibfnamefont{T.}~\bibnamefont{Fischer}},
  \bibinfo{author}{\bibfnamefont{S.}~\bibnamefont{Chakraborty}},
  \bibinfo{author}{\bibfnamefont{M.}~\bibnamefont{Giannotti}},
  \bibinfo{author}{\bibfnamefont{A.}~\bibnamefont{Mirizzi}},
  \bibinfo{author}{\bibfnamefont{A.}~\bibnamefont{Payez}}, \bibnamefont{and}
  \bibinfo{author}{\bibfnamefont{A.}~\bibnamefont{Ringwald}},
  \bibinfo{journal}{Phys. Rev. D} \textbf{\bibinfo{volume}{94}},
  \bibinfo{pages}{085012} (\bibinfo{year}{2016}), \eprint{1605.08780}.

\bibitem[{\citenamefont{Carenza et~al.}(2019)\citenamefont{Carenza, Fischer,
  Giannotti, Guo, Mart\'\i{}nez-Pinedo, and Mirizzi}}]{Carenza:2019pxu}
\bibinfo{author}{\bibfnamefont{P.}~\bibnamefont{Carenza}},
  \bibinfo{author}{\bibfnamefont{T.}~\bibnamefont{Fischer}},
  \bibinfo{author}{\bibfnamefont{M.}~\bibnamefont{Giannotti}},
  \bibinfo{author}{\bibfnamefont{G.}~\bibnamefont{Guo}},
  \bibinfo{author}{\bibfnamefont{G.}~\bibnamefont{Mart\'\i{}nez-Pinedo}},
  \bibnamefont{and} \bibinfo{author}{\bibfnamefont{A.}~\bibnamefont{Mirizzi}},
  \bibinfo{journal}{JCAP} \textbf{\bibinfo{volume}{10}}, \bibinfo{pages}{016}
  (\bibinfo{year}{2019}), \bibinfo{note}{[Erratum: JCAP 05, E01 (2020)]},
  \eprint{1906.11844}.

\bibitem[{\citenamefont{Fore and Reddy}(2020)}]{Fore:2019wib}
\bibinfo{author}{\bibfnamefont{B.}~\bibnamefont{Fore}} \bibnamefont{and}
  \bibinfo{author}{\bibfnamefont{S.}~\bibnamefont{Reddy}},
  \bibinfo{journal}{Phys. Rev. C} \textbf{\bibinfo{volume}{101}},
  \bibinfo{pages}{035809} (\bibinfo{year}{2020}), \eprint{1911.02632}.

\bibitem[{\citenamefont{Burrows et~al.}(1989)\citenamefont{Burrows, Turner, and
  Brinkmann}}]{Burrows:1988ah}
\bibinfo{author}{\bibfnamefont{A.}~\bibnamefont{Burrows}},
  \bibinfo{author}{\bibfnamefont{M.~S.} \bibnamefont{Turner}},
  \bibnamefont{and}
  \bibinfo{author}{\bibfnamefont{R.}~\bibnamefont{Brinkmann}},
  \bibinfo{journal}{Phys. Rev. D} \textbf{\bibinfo{volume}{39}},
  \bibinfo{pages}{1020} (\bibinfo{year}{1989}).

\bibitem[{\citenamefont{Burrows et~al.}(1990)\citenamefont{Burrows, Ressell,
  and Turner}}]{Burrows:1990pk}
\bibinfo{author}{\bibfnamefont{A.}~\bibnamefont{Burrows}},
  \bibinfo{author}{\bibfnamefont{M.}~\bibnamefont{Ressell}}, \bibnamefont{and}
  \bibinfo{author}{\bibfnamefont{M.~S.} \bibnamefont{Turner}},
  \bibinfo{journal}{Phys. Rev. D} \textbf{\bibinfo{volume}{42}},
  \bibinfo{pages}{3297} (\bibinfo{year}{1990}).

\bibitem[{\citenamefont{Raffelt and Seckel}(1991)}]{Raffelt:1991pw}
\bibinfo{author}{\bibfnamefont{G.}~\bibnamefont{Raffelt}} \bibnamefont{and}
  \bibinfo{author}{\bibfnamefont{D.}~\bibnamefont{Seckel}},
  \bibinfo{journal}{Phys. Rev. Lett.} \textbf{\bibinfo{volume}{67}},
  \bibinfo{pages}{2605} (\bibinfo{year}{1991}).

\bibitem[{\citenamefont{Hannestad and Raffelt}(1998)}]{Hannestad:1997gc}
\bibinfo{author}{\bibfnamefont{S.}~\bibnamefont{Hannestad}} \bibnamefont{and}
  \bibinfo{author}{\bibfnamefont{G.}~\bibnamefont{Raffelt}},
  \bibinfo{journal}{Astrophys. J.} \textbf{\bibinfo{volume}{507}},
  \bibinfo{pages}{339} (\bibinfo{year}{1998}), \eprint{astro-ph/9711132}.

\bibitem[{\citenamefont{Raffelt and Strobel}(1997)}]{Raffelt:1996di}
\bibinfo{author}{\bibfnamefont{G.}~\bibnamefont{Raffelt}} \bibnamefont{and}
  \bibinfo{author}{\bibfnamefont{T.}~\bibnamefont{Strobel}},
  \bibinfo{journal}{Phys. Rev. D} \textbf{\bibinfo{volume}{55}},
  \bibinfo{pages}{523} (\bibinfo{year}{1997}), \eprint{astro-ph/9610193}.

\bibitem[{\citenamefont{Migdal et~al.}(1990)\citenamefont{Migdal, Saperstein,
  Troitsky, and Voskresensky}}]{Migdal:1990vm}
\bibinfo{author}{\bibfnamefont{A.~B.} \bibnamefont{Migdal}},
  \bibinfo{author}{\bibfnamefont{E.}~\bibnamefont{Saperstein}},
  \bibinfo{author}{\bibfnamefont{M.}~\bibnamefont{Troitsky}}, \bibnamefont{and}
  \bibinfo{author}{\bibfnamefont{D.}~\bibnamefont{Voskresensky}},
  \bibinfo{journal}{Phys. Rept.} \textbf{\bibinfo{volume}{192}},
  \bibinfo{pages}{179} (\bibinfo{year}{1990}).

\bibitem[{\citenamefont{Tamborra et~al.}(2017)\citenamefont{Tamborra,
  Huedepohl, Raffelt, and Janka}}]{Tamborra:2017ubu}
\bibinfo{author}{\bibfnamefont{I.}~\bibnamefont{Tamborra}},
  \bibinfo{author}{\bibfnamefont{L.}~\bibnamefont{Huedepohl}},
  \bibinfo{author}{\bibfnamefont{G.}~\bibnamefont{Raffelt}}, \bibnamefont{and}
  \bibinfo{author}{\bibfnamefont{H.-T.} \bibnamefont{Janka}},
  \bibinfo{journal}{Astrophys. J.} \textbf{\bibinfo{volume}{839}},
  \bibinfo{pages}{132} (\bibinfo{year}{2017}), \eprint{1702.00060}.

\bibitem[{\citenamefont{Kim}(1979)}]{Kim:1979if}
\bibinfo{author}{\bibfnamefont{J.~E.} \bibnamefont{Kim}},
  \bibinfo{journal}{Phys. Rev. Lett.} \textbf{\bibinfo{volume}{43}},
  \bibinfo{pages}{103} (\bibinfo{year}{1979}).

\bibitem[{\citenamefont{Shifman et~al.}(1980)\citenamefont{Shifman, Vainshtein,
  and Zakharov}}]{Shifman:1979if}
\bibinfo{author}{\bibfnamefont{M.~A.} \bibnamefont{Shifman}},
  \bibinfo{author}{\bibfnamefont{A.}~\bibnamefont{Vainshtein}},
  \bibnamefont{and} \bibinfo{author}{\bibfnamefont{V.~I.}
  \bibnamefont{Zakharov}}, \bibinfo{journal}{Nucl. Phys. B}
  \textbf{\bibinfo{volume}{166}}, \bibinfo{pages}{493} (\bibinfo{year}{1980}).

\bibitem[{\citenamefont{Zhitnitsky}(1980)}]{Zhitnitsky:1980tq}
\bibinfo{author}{\bibfnamefont{A.}~\bibnamefont{Zhitnitsky}},
  \bibinfo{journal}{Sov. J. Nucl. Phys.} \textbf{\bibinfo{volume}{31}},
  \bibinfo{pages}{260} (\bibinfo{year}{1980}).

\bibitem[{\citenamefont{Dine et~al.}(1981)\citenamefont{Dine, Fischler, and
  Srednicki}}]{Dine:1981rt}
\bibinfo{author}{\bibfnamefont{M.}~\bibnamefont{Dine}},
  \bibinfo{author}{\bibfnamefont{W.}~\bibnamefont{Fischler}}, \bibnamefont{and}
  \bibinfo{author}{\bibfnamefont{M.}~\bibnamefont{Srednicki}},
  \bibinfo{journal}{Phys. Lett. B} \textbf{\bibinfo{volume}{104}},
  \bibinfo{pages}{199} (\bibinfo{year}{1981}).

\bibitem[{\citenamefont{Grilli~di Cortona et~al.}(2016)\citenamefont{Grilli~di
  Cortona, Hardy, Pardo~Vega, and Villadoro}}]{diCortona:2015ldu}
\bibinfo{author}{\bibfnamefont{G.}~\bibnamefont{Grilli~di Cortona}},
  \bibinfo{author}{\bibfnamefont{E.}~\bibnamefont{Hardy}},
  \bibinfo{author}{\bibfnamefont{J.}~\bibnamefont{Pardo~Vega}},
  \bibnamefont{and}
  \bibinfo{author}{\bibfnamefont{G.}~\bibnamefont{Villadoro}},
  \bibinfo{journal}{JHEP} \textbf{\bibinfo{volume}{01}}, \bibinfo{pages}{034}
  (\bibinfo{year}{2016}), \eprint{1511.02867}.

\bibitem[{\citenamefont{Di~Luzio
  et~al.}(2020{\natexlab{b}})\citenamefont{Di~Luzio, Giannotti, Nardi, and
  Visinelli}}]{DiLuzio:2020wdo}
\bibinfo{author}{\bibfnamefont{L.}~\bibnamefont{Di~Luzio}},
  \bibinfo{author}{\bibfnamefont{M.}~\bibnamefont{Giannotti}},
  \bibinfo{author}{\bibfnamefont{E.}~\bibnamefont{Nardi}}, \bibnamefont{and}
  \bibinfo{author}{\bibfnamefont{L.}~\bibnamefont{Visinelli}},
  \bibinfo{journal}{Phys. Rept.} \textbf{\bibinfo{volume}{870}},
  \bibinfo{pages}{1} (\bibinfo{year}{2020}{\natexlab{b}}), \eprint{2003.01100}.

\bibitem[{\citenamefont{Bartl et~al.}(2014)\citenamefont{Bartl, Pethick, and
  Schwenk}}]{Bartl:2014hoa}
\bibinfo{author}{\bibfnamefont{A.}~\bibnamefont{Bartl}},
  \bibinfo{author}{\bibfnamefont{C.}~\bibnamefont{Pethick}}, \bibnamefont{and}
  \bibinfo{author}{\bibfnamefont{A.}~\bibnamefont{Schwenk}},
  \bibinfo{journal}{Phys. Rev. Lett.} \textbf{\bibinfo{volume}{113}},
  \bibinfo{pages}{081101} (\bibinfo{year}{2014}), \eprint{1403.4114}.

\bibitem[{\citenamefont{Raffelt}(2008)}]{Raffelt:2006cw}
\bibinfo{author}{\bibfnamefont{G.~G.} \bibnamefont{Raffelt}},
  \bibinfo{journal}{Lect. Notes Phys.} \textbf{\bibinfo{volume}{741}},
  \bibinfo{pages}{51} (\bibinfo{year}{2008}), \eprint{hep-ph/0611350}.

\bibitem[{\citenamefont{Pons et~al.}(2000)\citenamefont{Pons, Reddy, Ellis,
  Prakash, and Lattimer}}]{Pons:2000iy}
\bibinfo{author}{\bibfnamefont{J.~A.} \bibnamefont{Pons}},
  \bibinfo{author}{\bibfnamefont{S.}~\bibnamefont{Reddy}},
  \bibinfo{author}{\bibfnamefont{P.~J.} \bibnamefont{Ellis}},
  \bibinfo{author}{\bibfnamefont{M.}~\bibnamefont{Prakash}}, \bibnamefont{and}
  \bibinfo{author}{\bibfnamefont{J.~M.} \bibnamefont{Lattimer}},
  \bibinfo{journal}{Phys. Rev. C} \textbf{\bibinfo{volume}{62}},
  \bibinfo{pages}{035803} (\bibinfo{year}{2000}), \eprint{nucl-th/0003008}.

\end{thebibliography}
\end{document}